\pdfoutput=1
\documentclass[prl,nofootinbib,twocolumn,floatfix,preprintnumbers,showpacs]{revtex4}

\usepackage[T1]{fontenc}
\usepackage[utf8]{inputenc}
\usepackage{lmodern}

\usepackage{verbatim}
\usepackage{natbib}
\usepackage[dvipsnames, usenames]{xcolor}
\definecolor{linkcolor}{rgb}{0.0,0.3,0.5}
\usepackage[hypertexnames=false, unicode, colorlinks=true, linkcolor=linkcolor,
citecolor=linkcolor, filecolor=linkcolor,urlcolor=linkcolor,
pdfusetitle]{hyperref}

\usepackage[all]{hypcap}
\usepackage{graphicx}
\usepackage{xspace}
\usepackage{amssymb}
\usepackage[normalem]{ulem} 
\usepackage{bm} 

\usepackage{microtype}

\usepackage[english]{babel}
\usepackage{blindtext}

\graphicspath{%
  {figs/}%
}

\DeclareMathAlphabet{\mathpzc}{OT1}{pzc}{m}{it}

\newcommand{\be}{\begin{equation}}
\newcommand{\ee}{\end{equation}}
\newcommand{\bea}{\begin{eqnarray}}
\newcommand{\eea}{\end{eqnarray}}

\newcommand{\nn}{\nonumber}

\begin{document}

\title{Impact of trans-Planckian quantum noise on the Primordial Gravitational Wave spectrum}

\newcommand{\INFN}{\affiliation{INFN - Sezione di Napoli, Complesso Univ. Monte S. Angelo, I-80126 Napoli, Italy}}

\newcommand{\UNI}{\affiliation{Dipartimento di Fisica ``Ettore Pancini”, Università degli studi di Napoli ``Federico II”,
Complesso Univ. Monte S. Angelo, I-80126 Napoli, Italy}}

\author{Mattia Cielo}
\email{mattia.cielo@unina.it}
\UNI
\INFN

\author{Gianpiero Mangano}
\email{gmangano@na.infn.it}
\UNI
\INFN

\author{Ofelia Pisanti}
\email{pisanti@na.infn.it}
\UNI
\INFN

\hypersetup{pdfauthor={Varma et al.}}

\date{\today}


\begin{abstract} 
We investigate the impact of stochastic quantum noise due to trans--Planckian effects on the primordial power spectrum for gravity waves during inflation. Given an energy scale $\Lambda$, expected to be close to the Planck scale $m_{Pl}$ and larger than the Hubble scale $H$, this noise is described in terms of a source term in the evolution equation for comoving modes $k$ which changes its amplitude growth from early times as long as the mode physical wavelength is smaller than $\Lambda^{ -1}$. We model the source term as due to a gas of black holes in the trans--Planckian regime and the corresponding Hawking radiation. In fact, for energy scales larger than, or of the order of $\Lambda$, it is expected that trapped surfaces may form due to large energy densities. At later times the evolution then follows the standard sourceless evolution. We find that this mechanism still leads to a scale-invariant power spectrum of tensor perturbations, with an amplitude that depends upon the ratio $\Lambda/m_{Pl}$.

\end{abstract}

\maketitle

\section{Introduction}
Since its birth \cite{Guth:1980zm, Starobinsky:1979ty, Linde:1981mu}, cosmic inflation represents the most promising mechanism able to solve several puzzles of the very early universe cosmology such as the horizon problem for the CMB patches in the sky, flatness, monopole problems, etc.. It was then realized that it naturally provides a natural description for the formation of the primordial density and tensor perturbations.
While the features of scalar perturbations are measured very precisely in CMB and large-scale structures and are in excellent agreement with the inflationary model, maybe the present smoking gun of the inflationary paradigm is the production of primordial gravitational waves. It is a peculiar outcome of inflation and, most importantly, it provides information about a very early stage of the evolution of the Universe. This is the reason why this stochastic background is getting more and more attention with many hopes toward the future space-interferometers (see for example \cite{Abbott2021, Abbott2021a, Collaboration2015, Ricciardone:2016ddg, KAGRA:2013rdx, LISACosmologyWorkingGroup:2022jok, KAGRA:2021kbb}).

The issue we address in this paper is the choice of the vacuum state from which perturbations evolve. The standard choice is the  ``Euclidean vacuum'' or  Bunch-Davies (BD) vacuum. This vacuum has been widely accepted in the literature for many reasons. It is the state which minimizes the Hamiltonian of the theory, consistent with the uncertainty principle and with the instantaneous Minkowskian vacuum. It is also conjectured that the BD vacuum is an attractor solution for inflation \cite{Armendariz-Picon:2006saa, Kaloper:2018zgi}. Finally, such a choice leads to a scale-invariant power spectrum, as strongly preferred by observations. 

Nevertheless, one can object that using an asymptotic Minkowskian configuration of the Universe has no fundamental underlying reasons and is a too strong condition. Indeed,  even if the mode initial state is chosen as the lowest energy one (i.e. the particle state), for fields evolving in curved spacetimes a pure particle state can evolve into a mixed state because of the time dependence of the total Hamiltonian. 
Moreover, taking the state at an initial conformal time $\tau \rightarrow - \infty$, is equivalent to considering modes $k$ of zero wavelength and infinite energy. This means that mode evolution goes through a trans--Planckian regime \cite{Cai:2019hge, Martin:2000bv}, about which we have no robust theoretical clues. During that transition, departures from the standard treatment may arise and new phenomena can emerge e.g. from Generalizations of the Uncertainty Principle (GUP) \cite{ Martin:2000xs, Martin:2003kp, Kempf:2001fa, Ashoorioon:2004vm, Ashoorioon:2005ep, Kempf:1994su}. A possible way out consists of moving the ultraviolet scale down to a more comfortable cut-off energy, somewhere in the range between the Planck scale $m_{Pl}$ and the inflationary energy scale $H$, where classical General Relativity can be safely used (\cite{Danielsson:2002kx, Green:2022ovz, Lesgourgues:1996jc, Tanaka:2000jw, Mukhanov:2007zz, Das:2022hjp}).

Whether we assume a BD vacuum or some initial time for perturbation evolution, a few questions remain unsolved. What are the behaviour and dynamics of fluctuations in the trans--Planckian regime? What are the imprints of this early stage on their eventual amplitude at horizon crossing? Is inflation washing out their effects? 

In this paper, we suggest that the evolution of tensor perturbation in the high energy regime above some mass-energy scale $\Lambda$, can be effectively described by adding a {\it stochastic} source term with zero mean to the evolution equations, which is due to the interaction of modes with the underlying background of fluctuations due to quantum gravity nonlinear effects. The scale $\Lambda$, expected to be of the order of $m_{Pl}$ or below,  but larger than the Hubble scale $H$ during inflation, is a free parameter defined by the condition that quantum gravity features are not negligible above this scale. In particular, to model this background,  we consider a scenario that we call the ``Black Hole (BH) gas'', based on the idea that at trans--Planckian energy densities trapped surfaces can form producing an environment of BH's. The corresponding emitted Hawking radiation couples to the evolution of a given tensor mode, due to the nonlinear nature of gravitational interactions and thus, acts as a source in the evolution equation. 

We will describe the model in detail in the next section. The main result we obtain is that the resulting power spectrum is still scale invariant and, depending on the value of $\Lambda$, is smaller or larger than the standard result from very tiny amount up to large enhancement as $\Lambda$ varies in the range $[H, m_{Pl}]$. 

On general grounds, we think that this approach can be also extended to scalar perturbations, with similar results. Scalar perturbations however are more sensitive to the details of the considered inflationary model, while tensor modes only depend upon initial conditions, possible quantum effects, as suggested here, and the value of the (almost) constant Hubble scale during inflation. In the following, we will concentrate on tensor modes and come back to the scalar part in the last section, which contains our conclusions and outlooks. We use natural units $c=\hbar=k_B=1$.

\section{Evolving tensor perturbation from the trans--Planckian regime with a stochastic source }
We first shortly  review the main aspects of the standard formalism describing cosmological tensor perturbation evolution during inflation (\cite{Boyle:2005se, Caprini:2015tfa, Chung:2003wn, Starobinsky:1979ty, Mukhanov:1981xt, Guzzetti2016, Kodama1984, Lyth:1996im, Maggiore2018, Rubakov:1982df, Wang:2013zva}). We start with the second-order Action  
\bea
    S&=& -\frac{1}{16 \pi G}\sum_r \int d\tau\, d \textbf{k}\, \frac{a(\tau)^2}{2} \Big[ \hat{h}'^r_\textbf{k}(\tau)\hat{h}'^r_{-\textbf{k}}(\tau) \nn \\
    &-& k^2 \hat{h}^r_\textbf{k}(\tau) \hat{h}^r_{-\textbf{k}}(\tau) 
    + 32 \pi G a(\tau)^2 \hat{{\Pi}}^r_{\textbf{k}}(\tau) \hat{h}^r_{-\textbf{k}}(\tau)\Big],
\eea
with $\tau \in]-\infty,0]$ the conformal time, $a(\tau)$ the scale factor and $\hat{h}^r_{\textbf{k}} (\tau)$ the tensor perturbations quantum fields, which can be expanded in fundamental solutions as
\be
    \hat{h}^r_{\textbf{k}} (\tau) = h_k(\tau) \hat{a}^r_{\textbf{k}} + h^*_k(\tau) \hat{a}^{r \dag}_{-\textbf{k}},
    \ee
where $\hat{a}^r_{\textbf{k}}$ and $\hat{a}^{r \dag}_{-\textbf{k}}$ satisfies canonical commutation relations.
We recall that, while the quantum operator has to have the physical dimension of the inverse of the square of a mass ( $[\hat{h}_{\textbf{k}}] \approx [M]^{-2}$), the mode function $h_{k}$ behaves like $[M]^{-1/2}$ since $[\hat{a}_{\textbf{k}}] \approx [M]^{-3/2}$.  

The source term $\hat{{\Pi}}^r_{\textbf{k}}(\tau) $ vanishes at linear order in the standard case and thus the $h_k(\tau)$ evolution is dictated by a homogeneous differential equation. The main issue of this paper is to study the consequences of a stochastic source in a model which will be detailed in the following. Taking for the moment  $\hat{{\Pi}}^r_{\textbf{k}}(\tau) =0$,
variation of the action with respect to $h_k$ leads to \be
h''_k + 2 \mathcal{H} h'_k + k^2 h_k = 0,
\label{omogenea}
\ee
and defining rescaled fields as 
\be
    v_{k}(\tau) = a(\tau)\, h_k(\tau),
\ee
we  obtain the well-known Mukhanov-Sasaki equation \cite{Mukhanov:1990me} 
\be
    v''_{k} + \Big( k^2 - \frac{a''}{a} \Big) v_k =0.
\ee
 This equation is solved by a linear combination of the Hankel functions weighted by the Bogoliubov coefficients $A_k$ and $B_k$
\be
h_{k}(\tau) = \frac{A_{k}}{a(\tau)} \frac{e^{- i k \tau}}{\sqrt{2k}} \Big(1 - \frac{i}{k \tau}\Big) + \frac{B_{k}}{a(\tau)} \frac{e^{ i k \tau}}{\sqrt{2k}} \Big(1 + \frac{i}{k \tau}\Big) \label{e:homsol}.
\ee
The coefficients $A_k$ and $B_k$ encode the natural ambiguity of the vacuum state in curved space-time due to the lack of timelike Killing vectors, and in order to characterize the solution one has to impose the initial condition for the mode functions. The standard choice is the  ``Euclidean vacuum'' or  Bunch-Davies vacuum with $A_k = 1$ and $B_k = 0$. 
which leads to a standard scale-invariant power spectrum, in agreement with observations
\bea
    P^t_{BD} &=& (64 \pi G) \frac{k^3}{2 \pi^2} \langle 0 | \hat{h}^\dag_{\textbf{k}}  \hat{h}_{\textbf{k}} |0\rangle \nn \\
    &=& (64 \pi G) \frac{k^3}{2 \pi^2}  |h_k|_{BD}^2 = \frac{16 H^2}{\pi m^2_{Pl}}.
\eea
Although the inflationary stage is not only a de Sitter space-time, since we only need a limited period with an exponential expansion, one can question whether using an asymptotic Minkowskian configuration of the Universe has fundamental underlying reasons and is a too strong condition. We stress again that taking the initial state at an initial time $\tau \rightarrow - \infty$, means that modes start at zero wavelength and with infinite energy. A possible way  out has been considered in \cite{Danielsson:2002kx} where the author  makes a different prescription introducing an initial time when the quantum fluctuations started evolving. This corresponds to recast creation and annihilation operators in terms of their value at a given time $\bar{\tau}_k$ \cite{Broy:2016zik}, such that the vacuum definition is
\be
\hat{a}_k(\bar{\tau}_k) |0, \bar{\tau}_k \rangle =0,
\ee
which in turn gives a relation between the Bogoliubov coefficients, $A_k$, and $B_k$. This is an example of what has been called $\alpha$ vacua (\cite{Allen:1985ux, Broy:2016zik, BouzariNezhad:2018zsi, Alberghi2003, Lemoine:2001ar, Agarwal:2012mq, Holman:2007na}). 

Imposing the vacuum at a given $\bar{\tau}_k$, introduces a physical cut-off scale, $m_{Pl} \geq \Lambda > H$,  such that the mode evolution begins only once $k=a\,\Lambda$. i.e. in quasi de Sitter \be
\bar{\tau}_k \simeq - \frac{\Lambda}{k H}. \label{e:taui}
\ee
In this case one obtains the following expression for the dimensionless power spectrum \cite{Broy:2016zik},
\be
P^t = \frac{16 H^2}{\pi m^2_{Pl}} \Bigl[
1 + \frac{H}{\Lambda}\sin \Bigl( 
\frac{2 \Lambda}{H}
\Bigr) + ...
\Bigr],
\ee
thus an oscillating feature that is subdominant with respect to the standard value as long as $\Lambda\gg H$.

In the following we will refer to the early trans--Planckian stage as that corresponding to mode energies larger than the scale $\Lambda$ above which quantum gravity effects are not negligible. $\Lambda$ is expected to be of the order of, or, possibly, smaller by a few orders of magnitudes than $m_{Pl}$. Since we have not yet a full theory of gravity in its quantum regime, though there are many ideas and suggestions about it (\cite{Esposito:2011rx, Kiefer:2007ria, Kiefer:2013jqa}), we can only list a few considerations which will bring us to the model we propose:
\begin{itemize}
\item[i)]  we expect linear approximation to be not an appropriate one when modes experience the high energy scale $\Lambda$ regime since nonlinear effects are crucial;
\item[ii)] once the perturbation wave number become sub--Planckian we can trust the standard linear evolution, i.e. the amplification till horizon crossing and the end of inflation. Yet, this evolution will keep the memory of the initial condition at the matching point given by $k/a= \Lambda$;
\item[iii)] at energies larger than $\Lambda$ fluctuations may give rise to trapped surfaces \cite{Penrose:1964wq}, which is to say they can produce a black hole environment. As gravitational interactions are nonlinear, tensor fluctuations with a high $k$ will interact with this environment so that the evolution can be effectively described in terms of a non-homogeneous differential equation with a source term.
\end{itemize}

These points led us to consider the following model. We introduce a non-vanishing anisotropic stress tensor that will encode information about the chaotic environment from which each mode has to go through when it starts evolving. We generalize eq. (\ref{omogenea}), by adding a source term acting on modes inside the horizon satisfying the condition $k/a>\Lambda$, or $\tau<  \bar{\tau}_k$. This translates into a two-stage evolution
\bea
h''_k + 2 \mathcal{H} h'_k + k^2 h_k = 16\pi G a^2 \, \Pi_k ~~~&& \tau < \bar{\tau}_k \label{e:source} \\
h''_k + 2 \mathcal{H} h'_k + k^2 h_k = 0 ~~~~~~~~~~~~~&& \tau > \bar{\tau}_k, \label{e:nosource} 
\eea
with the following matching conditions 
\bea
 \lim_{\tau \rightarrow \bar{\tau_k}^-} h_k (\tau) &=& \lim_{\tau \rightarrow \bar{\tau_k}^+}h_k (\tau) \label{matching0}\\
\lim_{\tau \rightarrow \bar{\tau_k}^-} h'_k(\tau) &=& \lim_{\tau \rightarrow \bar{\tau_k}^+} h'_k(\tau). \label{matching1} 
\eea
At $\tau \rightarrow -\infty$ we assume the mode to be excited from the vacuum with the minimal possible energy, as in the Bunch-Davis case. Of course, this condition may be chosen differently, corresponding to some $\alpha$ vacuum. Yet, we are interested here in how the trans--Planckian regime changes mode evolution with respect to the standard case, so in order to make this comparison it is appropriate to use the standard choice $A_k=1$, $B_k=0$ for $\tau\rightarrow -\infty$. The role of a different choice for the initial condition has not been pursued here and may deserve further studies.

Differently than in the approach  \cite{Broy:2016zik} described before,  $\bar{\tau}_k$ is now the time where the source term switch off and the evolution of the (now sub--Planckian) mode is the usual one.

We note that the shear source has to be quantized as the quantum metric fluctuations
\be
    \hat{\Pi}^r_{\textbf{k}} (\tau) = \Pi_k(\tau) \hat{a}^r_{\textbf{k}} + \Pi^*_k(\tau) \hat{a}^{r \dag}_{-\textbf{k}},
\ee
and we suppose that $\Pi_k$ is a stochastic incoherent source that satisfies the following relations ($\langle...\rangle$ denotes average over the probability distribution)
\bea
    \langle \Pi_k(\tau) \rangle &=& 0 \nn \\
    \langle \Pi_k (\tau)\Pi_k^* (\tau') \rangle &=& \Lambda^6 \delta (\tau - \tau') \left|F\left( \frac{k}{a \Lambda}, \frac{\Lambda}{m_{Pl}} \right)\right|^2.
\label{conditions}
\eea

 The prefactor accounts for the dimensionality of $\Pi_k$ and has been chosen as the correct power of the typical scale of the trans--Planckian phase, while $F$ is an adimensional function accounting for the dependence of the source $\Pi_k(\tau)$ on $k$ and on the relative value of $\Lambda$ with respect to the Planck scale. The other dimensional relevant parameter, the Hubble scale during inflation, is fixed to $H/m_{Pl} = 10^{-6}$.

Eq.s (\ref{e:source}) and (\ref{e:nosource}) have different solutions, which have to be matched at the transition time, $\bar{\tau}_k$. The homogeneous solution is given by eq. (\ref{e:homsol}), while in presence of the source ($\tau\leq\bar{\tau}(k)$) we have
\be
h_k (\tau) = \frac{16\pi G}{a(\tau)}\,\int^{\tau}_{-\infty} d \tau' a(\tau ') G_k(\tau, \tau') \Pi_k (\tau'),
\ee
in terms of the Green function, $G_k(\tau, \tau')$ (see \cite{Biagetti:2013kwa}), 
\bea
G_k(\tau, \tau') &=& \frac{e^{- i k(\tau + \tau')}}{2 k^3 \tau'^2} \Big[e^{2ik\tau}(1 - i k \tau)(-i + k \tau') \nn \\
&+& e^{2 i k \tau'}(1 + i k \tau)(i + k\tau') \Big] \Theta(\tau - \tau').
\eea
Using the matching conditions (\ref{matching0}), (\ref{matching1}), and making the average as in the Langevin approach to Brownian motion, we find the following expression for the Bogoliubov coefficients 
\bea
\frac{A_k}{a(\bar{\tau}_k)} &=& e^{i k \bar{\tau}_k}[ h(\bar{\tau}_k) (- 1 + ik \bar{\tau}_k + k^2 \bar{\tau}_k^2) \nonumber \\
&-& h'_k(\bar{\tau}_k)(\bar{\tau}_k - i k \bar{\tau}_k^2)]\left(\sqrt{2} k^{3/2} \bar{\tau}_k^2\right)^{-1} \nonumber\\
\frac{B_k}{a(\bar{\tau}_k)}&=& e^{-i k \bar{\tau}_k}[h(\bar{\tau}_k) (- 1 - ik \bar{\tau}_k + k^2 \bar{\tau}_k^2) \nonumber \\
&-& h'_k(\bar{\tau}_k)(\bar{\tau}_k + i k \bar{\tau}_k^2)] \left(\sqrt{2} k^{3/2} \bar{\tau}_k^2\right)^{-1}. 
\eea
Recalling that $h(\tau) \sim k^{-1/2}$ we see that  they depend on the adimensional quantity $k \bar{\tau}_k= -\Lambda/H$, so that the power spectrum $P^t$ is still predicted to be {\it scale invariant} (up to slow roll corrections). At the horizon crossing, $k \sim a H$ 
\bea
P^t &=& P_{BD}^t \Big[ 1+ |B_k|^2 \Big(2 + \frac{2 k \bar{\tau}_k +i}{i}e^{2ik\bar{\tau}_k} \nn \\
&-& \frac{2 k \bar{\tau}_k -i}{i}e^{-2ik\bar{\tau}_k} \Big) \Big],  \label{powerratio}
\eea
where $B_k$ can be easily found in terms of  integral functions of the source $\Pi_k$ and $P_{BD}^t$ represents the standard contribution from the BD vacuum. Their expressions are though, quite involved and lengthy, and  we do not report them explicitly for the sake of brevity.

A particular model for the shear source is specified by a choice of the adimensional function $F(k/(a \Lambda), \Lambda/m_{Pl})$. We consider here a scenario which we name the ``BH gas'' model. As we mentioned, during the trans--Planckian phase, modes with $k>\Lambda$ experience quantum gravity effects which we describe in terms of interactions with the BH gas which are formed when trapped surfaces create. This background produces particles, in particular gravitational waves, through Hawking radiation, which act as a source for $h(\tau)$. Given a probability distribution of the BH's as a function of their mass $M$, $\xi(M)$, and approximating the Hawking emission spectrum with a Boltzmann shape we have 
\be 
F\left( \frac{k}{a \Lambda}, \frac{\Lambda}{m_{Pl}} \right) = \int_0^\infty \xi(M) \exp\left( - \frac{k}{a} \frac{8 \pi M}{m_{Pl}^2}\right) dM.
\ee
The distribution $\xi(M)$ depends upon $\Lambda$ which represents the typical scale above which quantum effects are not negligible, and thus, a natural cut-off for black hole mass distribution. We use a simple exponential function, as we do not expect any particular feature in $\xi(M)$
\be \xi(M) \, dM= \frac{1}{\Lambda} e^{-M/\Lambda} \,dM,
\ee
and we finally obtain
\be F\left( \frac{k}{a \Lambda}, \frac{\Lambda}{m_{Pl}} \right) = \left( 1+ \frac{k}{a \Lambda} \frac{8 \pi \Lambda^2}{m_{Pl}^2} \right)^{-1}. \label{BHgas}
\ee
For $\Lambda \sim m_{Pl}$ the $k$-dependent term dominates since  $k/(a \Lambda) >1$, but it becomes less important as $\Lambda$ decreases since its contribution is significant in a narrower $\tau$ interval.

We have numerically solved eq.s (\ref{e:source},\ref{e:nosource}) using (\ref{BHgas}) and obtained the Bogoliubov coefficients $A_k$ and $B_k$.
 \begin{figure}
    \centering
    \includegraphics[width = 1.0\linewidth]{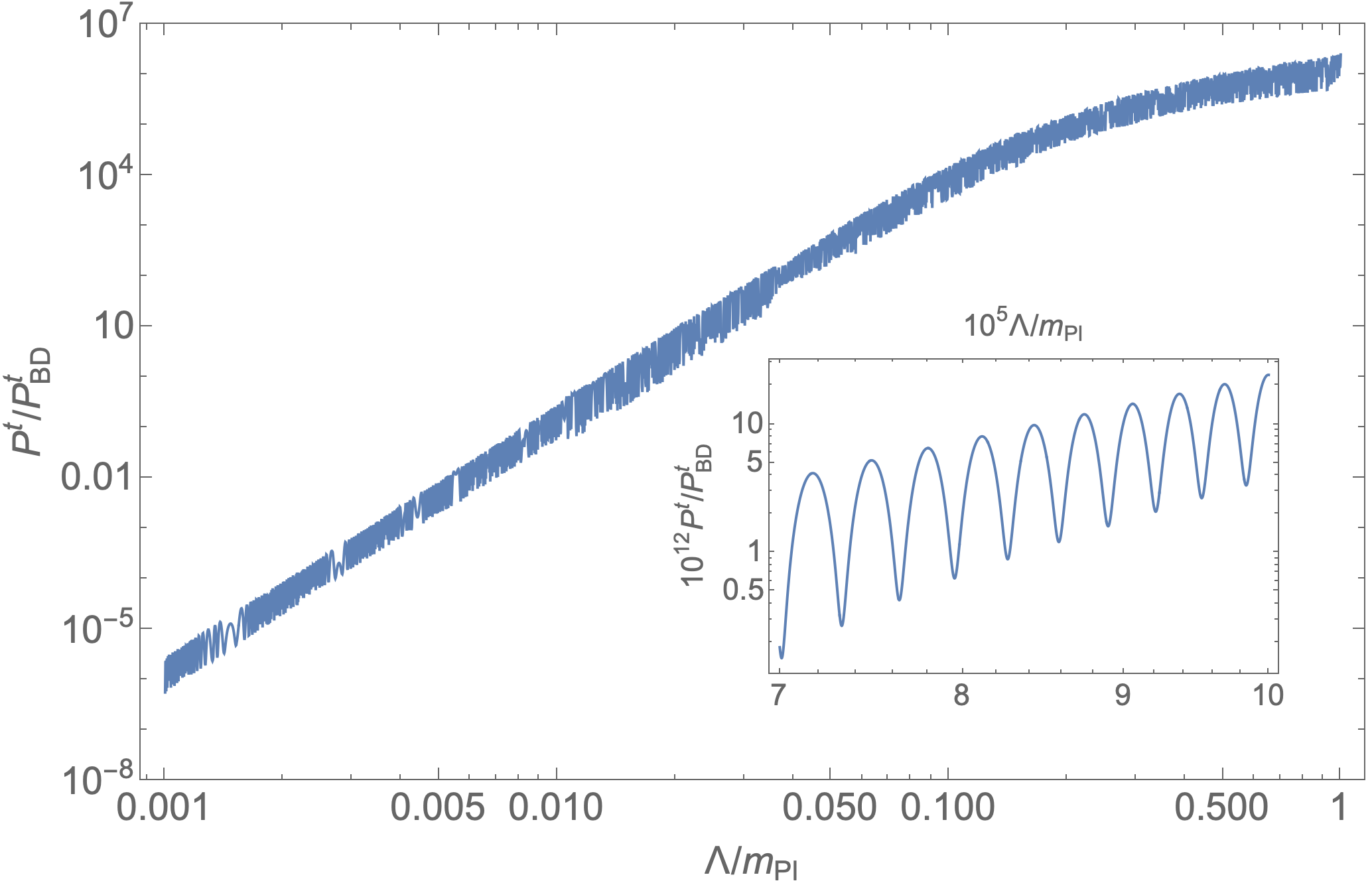}
    \caption{The scale-invariant tensor power spectrum normalized to the standard result $P^t_{BD}$ versus the ratio $\Lambda/m_{Pl}$. The zoom--in plot shows the oscillatory behaviour in a narrow range for $\Lambda/m_{Pl}$. See also the text.}
    \label{powersp}
 \end{figure}
The tensor power spectrum ratio $P^t/P^t_{BD}$ for a given $k$ (we recall that it is scale-invariant up to slow roll corrections as in the standard case) is shown in Fig. \ref{powersp} as a function of the ratio $\Lambda/m_{Pl}$. Two features should be noticed. First of all the very rapid oscillatory behaviour is due to the fact that $P^t$ depends upon oscillating trigonometric functions of the variable $\Lambda/H$ which in the selected range on the $x$-axis is very large (greater than $10^3$) since we have chosen $H/m_{Pl} \simeq 10^{-6}$. In the Figure is reported a zoom of the power spectrum ratio in a narrow range and for very small values of $\Lambda$ to appreciate more clearly this oscillatory behaviour. For higher and more natural values of $\Lambda$ the oscillations are so rapid that it is hard to clearly distinguish them. We also see from the plot that the value of $P^t/P^t_{BD}$ spans several orders of magnitude, from very small values up to a factor $10^6$  if $\Lambda= m_{Pl}$. This is due to two combining effects, the interference term in Eq. (\ref{powerratio}) which produces a rapid oscillating term in the evolution of perturbation, and the overall source scale $\Lambda^3$ which weights its amplitude. Notice that the predicted power spectrum is of the order of $P^t_{BD}$ for $\Lambda$ which is only two orders of magnitude smaller than the Planck scale, which is quite reasonable, meaning that for such a scale we reconstruct a scale-invariant spectrum of tensor perturbation with the correct amplitude. Might $\Lambda$ be larger or smaller?

\section{Conclusions and Outlooks}

The picture of how perturbations are produced during inflation is at the same time simple and successful. It is simple because it relies on the very distinctive feature of inflation, i.e. a quasi de Sitter stage during which the event horizon is almost constant. It is successful because this feature naturally provides a scale invariant spectrum, in agreement with observations. What remains, perhaps, to be understood is the effects of the early stages of perturbation evolution, when they undergo a trans--Planckian regime where quantum gravity features may not be negligible and the linear approximation may fail. This problem can be translated into the question of what is the initial condition for perturbation modes when they become ``classical'' from the point of view of gravitational interactions, i.e. when the wavenumber $k$ is smaller than $a \Lambda$. 

In this paper, we have considered the possibility that this initial condition, differently from the BD assumption of an asymptotic Minkowskian vacuum, is the outcome of an earlier stage where tensor perturbations are sourced by a shear term due to interactions with a stochastic background of excitations due to quantum gravity effects. In particular, we modelled these excitations as produced by Hawking radiation from a ``BH gas'', since at high trans--Planckian energies trapped surfaces and BH excitations may form.

Our result is that this model still predicts a scale invariant spectrum for tensor perturbation with an amplitude which grows with the scale $\Lambda$ from tiny values, for the unrealistic case of $\Lambda\sim H$ and due to interference effects, to a large enhancement for $\Lambda\sim m_{Pl}$. For $\Lambda/m_{Pl}\sim 10^{-2}$ it agrees with the $BD$ result. 

On the other hand, we know that the tensor to scalar perturbation ratio is strongly constrained by CMB data, in particular by the Planck Collaboration results \cite{Planck:2018jri}. Our point is that scalar perturbations too, would experience the same behaviour in the trans--Planckian regime we have described so far, and thus, we expect at a first glance that  the tensor to scalar ratio is {\it independent} from $\Lambda$. This is currently under study  
\cite{Cielo}.
We notice that in the standard case scalar perturbations depend only on the Hubble parameter and the features of the inflaton potential, which in fact can be constrained in the relevant e-fold region \cite{Planck:2018jri}. If the parameter $\Lambda$ would be added to the dynamics of fluctuations in the high energy regime, this may provide different results for what we know about the inflation dynamics, as we expect some extra degeneracies among the slow-roll parameters and $\Lambda$, once the perturbation amplitude at large CMB scale is fixed by data.  Last but not least it might be also interesting to see the effects on non-gaussian features of the primordial spectrum in the scenario we have discussed \cite{Brahma:2013rua, CieloFasiello}.

\section{Acknowledgements}
Work supported by the Italian grant 2017W4HA7S “NAT-NET: Neutrino and Astroparticle Theory Network” (PRIN
2017) funded by the Italian Ministero dell’Istruzione, dell’Universit\'a e della Ricerca (MIUR), and Iniziativa Specifica TAsP of INFN. M.C. thanks Matteo Fasiello for enlightening discussion. 

\bibliography{PGWsbib.bib}

\begin{thebibliography}{53}
\expandafter\ifx\csname natexlab\endcsname\relax\def\natexlab#1{#1}\fi
\expandafter\ifx\csname bibnamefont\endcsname\relax
  \def\bibnamefont#1{#1}\fi
\expandafter\ifx\csname bibfnamefont\endcsname\relax
  \def\bibfnamefont#1{#1}\fi
\expandafter\ifx\csname citenamefont\endcsname\relax
  \def\citenamefont#1{#1}\fi
\expandafter\ifx\csname url\endcsname\relax
  \def\url#1{\texttt{#1}}\fi
\expandafter\ifx\csname urlprefix\endcsname\relax\def\urlprefix{URL }\fi
\providecommand{\bibinfo}[2]{#2}
\providecommand{\eprint}[2][]{\url{#2}}

\bibitem[{\citenamefont{Guth}(1981)}]{Guth:1980zm}
\bibinfo{author}{\bibfnamefont{A.~H.} \bibnamefont{Guth}},
  \bibinfo{journal}{Phys. Rev. D} \textbf{\bibinfo{volume}{23}},
  \bibinfo{pages}{347} (\bibinfo{year}{1981}).

\bibitem[{\citenamefont{Starobinsky}(1979)}]{Starobinsky:1979ty}
\bibinfo{author}{\bibfnamefont{A.~A.} \bibnamefont{Starobinsky}},
  \bibinfo{journal}{JETP Lett.} \textbf{\bibinfo{volume}{30}},
  \bibinfo{pages}{682} (\bibinfo{year}{1979}).

\bibitem[{\citenamefont{Linde}(1982)}]{Linde:1981mu}
\bibinfo{author}{\bibfnamefont{A.~D.} \bibnamefont{Linde}},
  \bibinfo{journal}{Phys. Lett. B} \textbf{\bibinfo{volume}{108}},
  \bibinfo{pages}{389} (\bibinfo{year}{1982}).

\bibitem[{\citenamefont{Abbott et~al.}(2021{\natexlab{a}})\citenamefont{Abbott,
  Abbott, Abbott, Abraham, Acernese, Ackley, Adams, Adhikari, Adya, Affeldt
  et~al.}}]{Abbott2021}
\bibinfo{author}{\bibfnamefont{B.~P.} \bibnamefont{Abbott}},
  \bibinfo{author}{\bibfnamefont{R.}~\bibnamefont{Abbott}},
  \bibinfo{author}{\bibfnamefont{T.~D.} \bibnamefont{Abbott}},
  \bibinfo{author}{\bibfnamefont{S.}~\bibnamefont{Abraham}},
  \bibinfo{author}{\bibfnamefont{F.}~\bibnamefont{Acernese}},
  \bibinfo{author}{\bibfnamefont{K.}~\bibnamefont{Ackley}},
  \bibinfo{author}{\bibfnamefont{C.}~\bibnamefont{Adams}},
  \bibinfo{author}{\bibfnamefont{R.~X.} \bibnamefont{Adhikari}},
  \bibinfo{author}{\bibfnamefont{V.~B.} \bibnamefont{Adya}},
  \bibinfo{author}{\bibfnamefont{C.}~\bibnamefont{Affeldt}},
  \bibnamefont{et~al.}, \textbf{\bibinfo{volume}{9}}, \bibinfo{pages}{73}
  (\bibinfo{year}{2021}{\natexlab{a}}), \eprint{1908.06060v5}.

\bibitem[{\citenamefont{Abbott et~al.}(2021{\natexlab{b}})\citenamefont{Abbott,
  Abe, Acernese, Ackley, Adhikari, Adhikari, Adkins, Adya, Affeldt, Agarwal
  et~al.}}]{Abbott2021a}
\bibinfo{author}{\bibfnamefont{R.}~\bibnamefont{Abbott}},
  \bibinfo{author}{\bibfnamefont{H.}~\bibnamefont{Abe}},
  \bibinfo{author}{\bibfnamefont{F.}~\bibnamefont{Acernese}},
  \bibinfo{author}{\bibfnamefont{K.}~\bibnamefont{Ackley}},
  \bibinfo{author}{\bibfnamefont{N.}~\bibnamefont{Adhikari}},
  \bibinfo{author}{\bibfnamefont{R.~X.} \bibnamefont{Adhikari}},
  \bibinfo{author}{\bibfnamefont{V.~K.} \bibnamefont{Adkins}},
  \bibinfo{author}{\bibfnamefont{V.~B.} \bibnamefont{Adya}},
  \bibinfo{author}{\bibfnamefont{C.}~\bibnamefont{Affeldt}},
  \bibinfo{author}{\bibfnamefont{D.}~\bibnamefont{Agarwal}},
  \bibnamefont{et~al.}, \textbf{\bibinfo{volume}{95}}, \bibinfo{pages}{76}
  (\bibinfo{year}{2021}{\natexlab{b}}), \eprint{2111.03604v2}.

\bibitem[{\citenamefont{Collaboration et~al.}(2015)\citenamefont{Collaboration,
  Ade, Aghanim, Armitage-Caplan, Arnaud, Ashdown, Atrio-Barandela, Aumont,
  Baccigalupi, Banday et~al.}}]{Collaboration2015}
\bibinfo{author}{\bibfnamefont{P.}~\bibnamefont{Collaboration}},
  \bibinfo{author}{\bibfnamefont{P.~A.~R.} \bibnamefont{Ade}},
  \bibinfo{author}{\bibfnamefont{N.}~\bibnamefont{Aghanim}},
  \bibinfo{author}{\bibfnamefont{C.}~\bibnamefont{Armitage-Caplan}},
  \bibinfo{author}{\bibfnamefont{M.}~\bibnamefont{Arnaud}},
  \bibinfo{author}{\bibfnamefont{M.}~\bibnamefont{Ashdown}},
  \bibinfo{author}{\bibfnamefont{F.}~\bibnamefont{Atrio-Barandela}},
  \bibinfo{author}{\bibfnamefont{J.}~\bibnamefont{Aumont}},
  \bibinfo{author}{\bibfnamefont{C.}~\bibnamefont{Baccigalupi}},
  \bibinfo{author}{\bibfnamefont{A.~J.} \bibnamefont{Banday}},
  \bibnamefont{et~al.}, \bibinfo{journal}{C. Hern{\'{a}}ndez-Monteagudo}
  \textbf{\bibinfo{volume}{26}}, \bibinfo{pages}{34} (\bibinfo{year}{2015}),
  \eprint{1303.5082v3}.

\bibitem[{\citenamefont{Ricciardone}(2017)}]{Ricciardone:2016ddg}
\bibinfo{author}{\bibfnamefont{A.}~\bibnamefont{Ricciardone}},
  \bibinfo{journal}{J. Phys. Conf. Ser.} \textbf{\bibinfo{volume}{840}},
  \bibinfo{pages}{012030} (\bibinfo{year}{2017}), \eprint{1612.06799}.

\bibitem[{\citenamefont{Abbott et~al.}(2018)}]{KAGRA:2013rdx}
\bibinfo{author}{\bibfnamefont{B.~P.} \bibnamefont{Abbott}}
  \bibnamefont{et~al.} (\bibinfo{collaboration}{KAGRA, LIGO Scientific, Virgo,
  VIRGO}), \bibinfo{journal}{Living Rev. Rel.} \textbf{\bibinfo{volume}{21}},
  \bibinfo{pages}{3} (\bibinfo{year}{2018}), \eprint{1304.0670}.

\bibitem[{\citenamefont{Auclair
  et~al.}(2022)}]{LISACosmologyWorkingGroup:2022jok}
\bibinfo{author}{\bibfnamefont{P.}~\bibnamefont{Auclair}} \bibnamefont{et~al.}
  (\bibinfo{collaboration}{LISA Cosmology Working Group})
  (\bibinfo{year}{2022}), \eprint{2204.05434}.

\bibitem[{\citenamefont{Abbott et~al.}(2021{\natexlab{c}})}]{KAGRA:2021kbb}
\bibinfo{author}{\bibfnamefont{R.}~\bibnamefont{Abbott}} \bibnamefont{et~al.}
  (\bibinfo{collaboration}{KAGRA, Virgo, LIGO Scientific}),
  \bibinfo{journal}{Phys. Rev. D} \textbf{\bibinfo{volume}{104}},
  \bibinfo{pages}{022004} (\bibinfo{year}{2021}{\natexlab{c}}),
  \eprint{2101.12130}.

\bibitem[{\citenamefont{Armendariz-Picon}(2007)}]{Armendariz-Picon:2006saa}
\bibinfo{author}{\bibfnamefont{C.}~\bibnamefont{Armendariz-Picon}},
  \bibinfo{journal}{JCAP} \textbf{\bibinfo{volume}{02}}, \bibinfo{pages}{031}
  (\bibinfo{year}{2007}), \eprint{astro-ph/0612288}.

\bibitem[{\citenamefont{Kaloper and Scargill}(2019)}]{Kaloper:2018zgi}
\bibinfo{author}{\bibfnamefont{N.}~\bibnamefont{Kaloper}} \bibnamefont{and}
  \bibinfo{author}{\bibfnamefont{J.}~\bibnamefont{Scargill}},
  \bibinfo{journal}{Phys. Rev. D} \textbf{\bibinfo{volume}{99}},
  \bibinfo{pages}{103514} (\bibinfo{year}{2019}), \eprint{1802.09554}.

\bibitem[{\citenamefont{Cai and Piao}(2020)}]{Cai:2019hge}
\bibinfo{author}{\bibfnamefont{Y.}~\bibnamefont{Cai}} \bibnamefont{and}
  \bibinfo{author}{\bibfnamefont{Y.-S.} \bibnamefont{Piao}},
  \bibinfo{journal}{Sci. China Phys. Mech. Astron.}
  \textbf{\bibinfo{volume}{63}}, \bibinfo{pages}{110411}
  (\bibinfo{year}{2020}), \eprint{1909.12719}.

\bibitem[{\citenamefont{Martin and Brandenberger}(2000)}]{Martin:2000bv}
\bibinfo{author}{\bibfnamefont{J.}~\bibnamefont{Martin}} \bibnamefont{and}
  \bibinfo{author}{\bibfnamefont{R.~H.} \bibnamefont{Brandenberger}}
  (\bibinfo{year}{2000}), pp. \bibinfo{pages}{2001--2002},
  \eprint{astro-ph/0012031}.

\bibitem[{\citenamefont{Martin and Brandenberger}(2001)}]{Martin:2000xs}
\bibinfo{author}{\bibfnamefont{J.}~\bibnamefont{Martin}} \bibnamefont{and}
  \bibinfo{author}{\bibfnamefont{R.~H.} \bibnamefont{Brandenberger}},
  \bibinfo{journal}{Phys. Rev. D} \textbf{\bibinfo{volume}{63}},
  \bibinfo{pages}{123501} (\bibinfo{year}{2001}), \eprint{hep-th/0005209}.

\bibitem[{\citenamefont{Martin and Brandenberger}(2003)}]{Martin:2003kp}
\bibinfo{author}{\bibfnamefont{J.}~\bibnamefont{Martin}} \bibnamefont{and}
  \bibinfo{author}{\bibfnamefont{R.}~\bibnamefont{Brandenberger}},
  \bibinfo{journal}{Phys. Rev. D} \textbf{\bibinfo{volume}{68}},
  \bibinfo{pages}{063513} (\bibinfo{year}{2003}), \eprint{hep-th/0305161}.

\bibitem[{\citenamefont{Kempf and Niemeyer}(2001)}]{Kempf:2001fa}
\bibinfo{author}{\bibfnamefont{A.}~\bibnamefont{Kempf}} \bibnamefont{and}
  \bibinfo{author}{\bibfnamefont{J.~C.} \bibnamefont{Niemeyer}},
  \bibinfo{journal}{Phys. Rev. D} \textbf{\bibinfo{volume}{64}},
  \bibinfo{pages}{103501} (\bibinfo{year}{2001}), \eprint{astro-ph/0103225}.

\bibitem[{\citenamefont{Ashoorioon
  et~al.}(2005{\natexlab{a}})\citenamefont{Ashoorioon, Kempf, and
  Mann}}]{Ashoorioon:2004vm}
\bibinfo{author}{\bibfnamefont{A.}~\bibnamefont{Ashoorioon}},
  \bibinfo{author}{\bibfnamefont{A.}~\bibnamefont{Kempf}}, \bibnamefont{and}
  \bibinfo{author}{\bibfnamefont{R.~B.} \bibnamefont{Mann}},
  \bibinfo{journal}{Phys. Rev. D} \textbf{\bibinfo{volume}{71}},
  \bibinfo{pages}{023503} (\bibinfo{year}{2005}{\natexlab{a}}),
  \eprint{astro-ph/0410139}.

\bibitem[{\citenamefont{Ashoorioon
  et~al.}(2005{\natexlab{b}})\citenamefont{Ashoorioon, Hovdebo, and
  Mann}}]{Ashoorioon:2005ep}
\bibinfo{author}{\bibfnamefont{A.}~\bibnamefont{Ashoorioon}},
  \bibinfo{author}{\bibfnamefont{J.~L.} \bibnamefont{Hovdebo}},
  \bibnamefont{and} \bibinfo{author}{\bibfnamefont{R.~B.} \bibnamefont{Mann}},
  \bibinfo{journal}{Nucl. Phys. B} \textbf{\bibinfo{volume}{727}},
  \bibinfo{pages}{63} (\bibinfo{year}{2005}{\natexlab{b}}),
  \eprint{gr-qc/0504135}.

\bibitem[{\citenamefont{Kempf et~al.}(1995)\citenamefont{Kempf, Mangano, and
  Mann}}]{Kempf:1994su}
\bibinfo{author}{\bibfnamefont{A.}~\bibnamefont{Kempf}},
  \bibinfo{author}{\bibfnamefont{G.}~\bibnamefont{Mangano}}, \bibnamefont{and}
  \bibinfo{author}{\bibfnamefont{R.~B.} \bibnamefont{Mann}},
  \bibinfo{journal}{Phys. Rev. D} \textbf{\bibinfo{volume}{52}},
  \bibinfo{pages}{1108} (\bibinfo{year}{1995}), \eprint{hep-th/9412167}.

\bibitem[{\citenamefont{Danielsson}(2002)}]{Danielsson:2002kx}
\bibinfo{author}{\bibfnamefont{U.~H.} \bibnamefont{Danielsson}},
  \bibinfo{journal}{Phys. Rev. D} \textbf{\bibinfo{volume}{66}},
  \bibinfo{pages}{023511} (\bibinfo{year}{2002}), \eprint{hep-th/0203198}.

\bibitem[{\citenamefont{Green}(2022)}]{Green:2022ovz}
\bibinfo{author}{\bibfnamefont{D.}~\bibnamefont{Green}} (\bibinfo{year}{2022}),
  \eprint{2210.05820}.

\bibitem[{\citenamefont{Lesgourgues et~al.}(1997)\citenamefont{Lesgourgues,
  Polarski, and Starobinsky}}]{Lesgourgues:1996jc}
\bibinfo{author}{\bibfnamefont{J.}~\bibnamefont{Lesgourgues}},
  \bibinfo{author}{\bibfnamefont{D.}~\bibnamefont{Polarski}}, \bibnamefont{and}
  \bibinfo{author}{\bibfnamefont{A.~A.} \bibnamefont{Starobinsky}},
  \bibinfo{journal}{Nucl. Phys. B} \textbf{\bibinfo{volume}{497}},
  \bibinfo{pages}{479} (\bibinfo{year}{1997}), \eprint{gr-qc/9611019}.

\bibitem[{\citenamefont{Tanaka}(2000)}]{Tanaka:2000jw}
\bibinfo{author}{\bibfnamefont{T.}~\bibnamefont{Tanaka}}
  (\bibinfo{year}{2000}), \eprint{astro-ph/0012431}.

\bibitem[{\citenamefont{Mukhanov and Winitzki}(2007)}]{Mukhanov:2007zz}
\bibinfo{author}{\bibfnamefont{V.}~\bibnamefont{Mukhanov}} \bibnamefont{and}
  \bibinfo{author}{\bibfnamefont{S.}~\bibnamefont{Winitzki}},
  \emph{\bibinfo{title}{{Introduction to quantum effects in gravity}}}
  (\bibinfo{publisher}{Cambridge University Press}, \bibinfo{year}{2007}).

\bibitem[{\citenamefont{Das et~al.}(2022)\citenamefont{Das, Shankaranarayanan,
  and Todorinov}}]{Das:2022hjp}
\bibinfo{author}{\bibfnamefont{S.}~\bibnamefont{Das}},
  \bibinfo{author}{\bibfnamefont{S.}~\bibnamefont{Shankaranarayanan}},
  \bibnamefont{and}
  \bibinfo{author}{\bibfnamefont{V.}~\bibnamefont{Todorinov}},
  \bibinfo{journal}{Phys. Lett. B} \textbf{\bibinfo{volume}{835}},
  \bibinfo{pages}{137511} (\bibinfo{year}{2022}), \eprint{2208.11095}.

\bibitem[{\citenamefont{Boyle and Steinhardt}(2008)}]{Boyle:2005se}
\bibinfo{author}{\bibfnamefont{L.~A.} \bibnamefont{Boyle}} \bibnamefont{and}
  \bibinfo{author}{\bibfnamefont{P.~J.} \bibnamefont{Steinhardt}},
  \bibinfo{journal}{Phys. Rev. D} \textbf{\bibinfo{volume}{77}},
  \bibinfo{pages}{063504} (\bibinfo{year}{2008}), \eprint{astro-ph/0512014}.

\bibitem[{\citenamefont{Caprini}(2015)}]{Caprini:2015tfa}
\bibinfo{author}{\bibfnamefont{C.}~\bibnamefont{Caprini}}, \bibinfo{journal}{J.
  Phys. Conf. Ser.} \textbf{\bibinfo{volume}{610}}, \bibinfo{pages}{012004}
  (\bibinfo{year}{2015}), \eprint{1501.01174}.

\bibitem[{\citenamefont{Chung et~al.}(2003)\citenamefont{Chung, Notari, and
  Riotto}}]{Chung:2003wn}
\bibinfo{author}{\bibfnamefont{D.~J.~H.} \bibnamefont{Chung}},
  \bibinfo{author}{\bibfnamefont{A.}~\bibnamefont{Notari}}, \bibnamefont{and}
  \bibinfo{author}{\bibfnamefont{A.}~\bibnamefont{Riotto}},
  \bibinfo{journal}{JCAP} \textbf{\bibinfo{volume}{10}}, \bibinfo{pages}{012}
  (\bibinfo{year}{2003}), \eprint{hep-ph/0305074}.

\bibitem[{\citenamefont{Mukhanov and Chibisov}(1981)}]{Mukhanov:1981xt}
\bibinfo{author}{\bibfnamefont{V.~F.} \bibnamefont{Mukhanov}} \bibnamefont{and}
  \bibinfo{author}{\bibfnamefont{G.~V.} \bibnamefont{Chibisov}},
  \bibinfo{journal}{JETP Lett.} \textbf{\bibinfo{volume}{33}},
  \bibinfo{pages}{532} (\bibinfo{year}{1981}).

\bibitem[{\citenamefont{Guzzetti et~al.}(2016)\citenamefont{Guzzetti, Bartolo,
  Liguori, and Matarrese}}]{Guzzetti2016}
\bibinfo{author}{\bibfnamefont{M.~C.} \bibnamefont{Guzzetti}},
  \bibinfo{author}{\bibfnamefont{N.}~\bibnamefont{Bartolo}},
  \bibinfo{author}{\bibfnamefont{M.}~\bibnamefont{Liguori}}, \bibnamefont{and}
  \bibinfo{author}{\bibfnamefont{S.}~\bibnamefont{Matarrese}}
  (\bibinfo{year}{2016}), \eprint{1605.01615v3}.

\bibitem[{\citenamefont{Kodama and Sasaki}(1984)}]{Kodama1984}
\bibinfo{author}{\bibfnamefont{H.}~\bibnamefont{Kodama}} \bibnamefont{and}
  \bibinfo{author}{\bibfnamefont{M.}~\bibnamefont{Sasaki}},
  \bibinfo{journal}{Progress of Theoretical Physics Supplement}
  \textbf{\bibinfo{volume}{78}} (\bibinfo{year}{1984}).

\bibitem[{\citenamefont{Lyth}(1997)}]{Lyth:1996im}
\bibinfo{author}{\bibfnamefont{D.~H.} \bibnamefont{Lyth}},
  \bibinfo{journal}{Phys. Rev. Lett.} \textbf{\bibinfo{volume}{78}},
  \bibinfo{pages}{1861} (\bibinfo{year}{1997}), \eprint{hep-ph/9606387}.

\bibitem[{\citenamefont{Maggiore}(2018)}]{Maggiore2018}
\bibinfo{author}{\bibfnamefont{M.}~\bibnamefont{Maggiore}},
  \bibinfo{journal}{Gravitational Waves: Volume 2: Astrophysics and Cosmology}
  pp. \bibinfo{pages}{1--820} (\bibinfo{year}{2018}).

\bibitem[{\citenamefont{Rubakov et~al.}(1982)\citenamefont{Rubakov, Sazhin, and
  Veryaskin}}]{Rubakov:1982df}
\bibinfo{author}{\bibfnamefont{V.~A.} \bibnamefont{Rubakov}},
  \bibinfo{author}{\bibfnamefont{M.~V.} \bibnamefont{Sazhin}},
  \bibnamefont{and} \bibinfo{author}{\bibfnamefont{A.~V.}
  \bibnamefont{Veryaskin}}, \bibinfo{journal}{Phys. Lett. B}
  \textbf{\bibinfo{volume}{115}}, \bibinfo{pages}{189} (\bibinfo{year}{1982}).

\bibitem[{\citenamefont{Wang}(2014)}]{Wang:2013zva}
\bibinfo{author}{\bibfnamefont{Y.}~\bibnamefont{Wang}},
  \bibinfo{journal}{Commun. Theor. Phys.} \textbf{\bibinfo{volume}{62}},
  \bibinfo{pages}{109} (\bibinfo{year}{2014}), \eprint{1303.1523}.

\bibitem[{\citenamefont{Mukhanov et~al.}(1992)\citenamefont{Mukhanov, Feldman,
  and Brandenberger}}]{Mukhanov:1990me}
\bibinfo{author}{\bibfnamefont{V.~F.} \bibnamefont{Mukhanov}},
  \bibinfo{author}{\bibfnamefont{H.~A.} \bibnamefont{Feldman}},
  \bibnamefont{and} \bibinfo{author}{\bibfnamefont{R.~H.}
  \bibnamefont{Brandenberger}}, \bibinfo{journal}{Phys. Rept.}
  \textbf{\bibinfo{volume}{215}}, \bibinfo{pages}{203} (\bibinfo{year}{1992}).

\bibitem[{\citenamefont{Broy}(2016)}]{Broy:2016zik}
\bibinfo{author}{\bibfnamefont{B.~J.} \bibnamefont{Broy}},
  \bibinfo{journal}{Phys. Rev. D} \textbf{\bibinfo{volume}{94}},
  \bibinfo{pages}{103508} (\bibinfo{year}{2016}), \bibinfo{note}{[Addendum:
  Phys.Rev.D 94, 109901 (2016)]}, \eprint{1609.03570}.

\bibitem[{\citenamefont{Allen}(1985)}]{Allen:1985ux}
\bibinfo{author}{\bibfnamefont{B.}~\bibnamefont{Allen}},
  \bibinfo{journal}{Phys. Rev. D} \textbf{\bibinfo{volume}{32}},
  \bibinfo{pages}{3136} (\bibinfo{year}{1985}).

\bibitem[{\citenamefont{Bouzari~Nezhad and
  Shojai}(2018)}]{BouzariNezhad:2018zsi}
\bibinfo{author}{\bibfnamefont{H.}~\bibnamefont{Bouzari~Nezhad}}
  \bibnamefont{and} \bibinfo{author}{\bibfnamefont{F.}~\bibnamefont{Shojai}},
  \bibinfo{journal}{Phys. Rev. D} \textbf{\bibinfo{volume}{98}},
  \bibinfo{pages}{063512} (\bibinfo{year}{2018}), \eprint{1802.05537}.

\bibitem[{\citenamefont{Alberghi et~al.}(2003)\citenamefont{Alberghi, Casadio,
  and Tronconi}}]{Alberghi2003}
\bibinfo{author}{\bibfnamefont{G.~L.} \bibnamefont{Alberghi}},
  \bibinfo{author}{\bibfnamefont{R.}~\bibnamefont{Casadio}}, \bibnamefont{and}
  \bibinfo{author}{\bibfnamefont{A.}~\bibnamefont{Tronconi}}
  (\bibinfo{year}{2003}).

\bibitem[{\citenamefont{Lemoine et~al.}(2002)\citenamefont{Lemoine, Lubo,
  Martin, and Uzan}}]{Lemoine:2001ar}
\bibinfo{author}{\bibfnamefont{M.}~\bibnamefont{Lemoine}},
  \bibinfo{author}{\bibfnamefont{M.}~\bibnamefont{Lubo}},
  \bibinfo{author}{\bibfnamefont{J.}~\bibnamefont{Martin}}, \bibnamefont{and}
  \bibinfo{author}{\bibfnamefont{J.-P.} \bibnamefont{Uzan}},
  \bibinfo{journal}{Phys. Rev. D} \textbf{\bibinfo{volume}{65}},
  \bibinfo{pages}{023510} (\bibinfo{year}{2002}), \eprint{hep-th/0109128}.

\bibitem[{\citenamefont{Agarwal et~al.}(2013)\citenamefont{Agarwal, Holman,
  Tolley, and Lin}}]{Agarwal:2012mq}
\bibinfo{author}{\bibfnamefont{N.}~\bibnamefont{Agarwal}},
  \bibinfo{author}{\bibfnamefont{R.}~\bibnamefont{Holman}},
  \bibinfo{author}{\bibfnamefont{A.~J.} \bibnamefont{Tolley}},
  \bibnamefont{and} \bibinfo{author}{\bibfnamefont{J.}~\bibnamefont{Lin}},
  \bibinfo{journal}{JHEP} \textbf{\bibinfo{volume}{05}}, \bibinfo{pages}{085}
  (\bibinfo{year}{2013}), \eprint{1212.1172}.

\bibitem[{\citenamefont{Holman and Tolley}(2008)}]{Holman:2007na}
\bibinfo{author}{\bibfnamefont{R.}~\bibnamefont{Holman}} \bibnamefont{and}
  \bibinfo{author}{\bibfnamefont{A.~J.} \bibnamefont{Tolley}},
  \bibinfo{journal}{JCAP} \textbf{\bibinfo{volume}{05}}, \bibinfo{pages}{001}
  (\bibinfo{year}{2008}), \eprint{0710.1302}.

\bibitem[{\citenamefont{Esposito}(2011)}]{Esposito:2011rx}
\bibinfo{author}{\bibfnamefont{G.}~\bibnamefont{Esposito}}
  (\bibinfo{year}{2011}), \eprint{1108.3269}.

\bibitem[{\citenamefont{Kiefer}(2007)}]{Kiefer:2007ria}
\bibinfo{author}{\bibfnamefont{C.}~\bibnamefont{Kiefer}},
  \emph{\bibinfo{title}{{Quantum Gravity}}} (\bibinfo{publisher}{Oxford
  University Press}, \bibinfo{address}{New York}, \bibinfo{year}{2007}),
  \bibinfo{edition}{2nd} ed., ISBN \bibinfo{isbn}{978-0-19-921252-1}.

\bibitem[{\citenamefont{Kiefer}(2013)}]{Kiefer:2013jqa}
\bibinfo{author}{\bibfnamefont{C.}~\bibnamefont{Kiefer}},
  \bibinfo{journal}{ISRN Math. Phys.} \textbf{\bibinfo{volume}{2013}},
  \bibinfo{pages}{509316} (\bibinfo{year}{2013}), \eprint{1401.3578}.

\bibitem[{\citenamefont{Penrose}(1965)}]{Penrose:1964wq}
\bibinfo{author}{\bibfnamefont{R.}~\bibnamefont{Penrose}},
  \bibinfo{journal}{Phys. Rev. Lett.} \textbf{\bibinfo{volume}{14}},
  \bibinfo{pages}{57} (\bibinfo{year}{1965}).

\bibitem[{\citenamefont{Biagetti et~al.}(2013)\citenamefont{Biagetti, Fasiello,
  and Riotto}}]{Biagetti:2013kwa}
\bibinfo{author}{\bibfnamefont{M.}~\bibnamefont{Biagetti}},
  \bibinfo{author}{\bibfnamefont{M.}~\bibnamefont{Fasiello}}, \bibnamefont{and}
  \bibinfo{author}{\bibfnamefont{A.}~\bibnamefont{Riotto}},
  \bibinfo{journal}{Phys. Rev. D} \textbf{\bibinfo{volume}{88}},
  \bibinfo{pages}{103518} (\bibinfo{year}{2013}), \eprint{1305.7241}.

\bibitem[{\citenamefont{Akrami et~al.}(2020)}]{Planck:2018jri}
\bibinfo{author}{\bibfnamefont{Y.}~\bibnamefont{Akrami}} \bibnamefont{et~al.}
  (\bibinfo{collaboration}{Planck}), \bibinfo{journal}{Astron. Astrophys.}
  \textbf{\bibinfo{volume}{641}}, \bibinfo{pages}{A10} (\bibinfo{year}{2020}),
  \eprint{1807.06211}.

\bibitem[{\citenamefont{Cielo et~al.}(in preparation)}]{Cielo}
\bibinfo{author}{\bibfnamefont{M.}~\bibnamefont{Cielo}} \bibnamefont{et~al.}
  (\bibinfo{year}{in preparation}).

\bibitem[{\citenamefont{Brahma et~al.}(2014)\citenamefont{Brahma, Nelson, and
  Shandera}}]{Brahma:2013rua}
\bibinfo{author}{\bibfnamefont{S.}~\bibnamefont{Brahma}},
  \bibinfo{author}{\bibfnamefont{E.}~\bibnamefont{Nelson}}, \bibnamefont{and}
  \bibinfo{author}{\bibfnamefont{S.}~\bibnamefont{Shandera}},
  \bibinfo{journal}{Phys. Rev. D} \textbf{\bibinfo{volume}{89}},
  \bibinfo{pages}{023507} (\bibinfo{year}{2014}), \eprint{1310.0471}.

\bibitem[{\citenamefont{Cielo and Fasiello}(in preparation)}]{CieloFasiello}
\bibinfo{author}{\bibfnamefont{M.}~\bibnamefont{Cielo}} \bibnamefont{and}
  \bibinfo{author}{\bibfnamefont{M.}~\bibnamefont{Fasiello}} (\bibinfo{year}{in
  preparation}).

\end{thebibliography}
\end{document}